\documentclass[%
print,
 amsmath,amssymb,
 aps,
]{revtex4}

\usepackage[sc]{mathpazo}
\usepackage{graphicx}
\usepackage{dcolumn}
\usepackage{bm}
\usepackage{dsfont}
\usepackage[landscape,papersize={297.1mm,210mm},left=1.9cm,right=1.6cm,top=2.7cm,bottom=2.8cm]{geometry}
\usepackage[                   
            pdfstartview=FitH,
            colorlinks, 
            pdfborder=001,   
            linkcolor=blue,
            anchorcolor=blue,
            citecolor=blue,
            urlcolor=blue
            ]{hyperref}
\usepackage{amsthm}
\makeatletter

\newcommand{\Rmnum}[1]{\expandafter\@slowromancap\romannumeral #1@}
\makeatother

\begin{document}
\title{Planar two-region  multi-partite maximally entangled states}

\author{Yanwen Liang}
\affiliation{College of Physics, Hebei Key Laboratory of Photophysics Research and Application, Hebei Normal University, Shijiazhuang 050024, China}

\author{Fengli Yan}
\email{flyan@hebtu.edu.cn}
\affiliation{College of Physics, Hebei Key Laboratory of Photophysics Research and Application, Hebei Normal University, Shijiazhuang 050024, China}

\author{Ting Gao}
\email{gaoting@hebtu.edu.cn}
\affiliation{School of Mathematical Sciences, Hebei Normal University, Shijiazhuang 050024, China}

\begin{abstract}
 In entanglement theory,  there are different methods to consider one state being more entangled than another. The  "maximally"
entangled states in a multipartite system can be defined from an axiomatic perspective. According to different  criteria for selection, there are many specific types of quantum maximally entangled states, such as absolutely maximally entangled state, planar maximally entangled state and so on. In this paper we propose a new type of maximally entangled states, the planar two-region multipartite maximally entangled state. The requirement condition of this maximally entangled state  is weak than that  of the absolutely maximally entangled state and different from that of  the planar maximally entangled state. We show that there are  the two-region four-partite maximally entangled states in   4-qubit and 7-qubit planar systems, although there is no  absolutely maximally entangled state in these systems.  It is proved  that there are the planar two-region four-partite maximally entangled states in both even particle quantum systems and odd particle quantum systems. Additionally, based on some  planar two-region four-partite maximally entangled states,  the new  planar two-region four-partite maximally entangled states are generated. We also provide some important examples of  the planar two-region  multi-partite maximally entangled states.\\

\end{abstract}


\maketitle

\section{Introduction}
Multipartite quantum entanglement \cite{Einstein, Werner, Horodecki}   is a  distinctive feature
of quantum mechanics and  responsible for numerous practical
tasks \cite{Nielsen}. So quantum entanglement  is   an essential constituent part of quantum
information and at the same time the subject of philosophical debates \cite{Horodecki}.
By far bipartite entangled states have been studied sufficiently, however the knowledge about multipartite states is far from perfect.
From both a theoretical and practical perspective, multipartite
entangled states are  a very important resource in
quantum information science.

It is well-known that the maximally
entangled states (MES) play a very  important role in practical
quantum information processing.  However, in fact  there are different
ways to consider one state to be more entangled than another. A method for defining
 "maximally"
entangled states in  multipartite systems is based on an axiomatic approach. By using this method
 Gisin and Bechmann-Pasquinucci  identify $n$-qubit Greenberger-Horne-Zeilinger (GHZ)
states as maximally entangled in $n$-qubit systems \cite{Gisin}.
The reason
for regarding these states as maximally entangled may be
that GHZ
states of $n$ qubits  maximally violate the Bell-Klyshko inequalities  \cite{Klyshko}. Moreover, Chen proved that GHZ states in $n$-qubit systems and the states obtained from them by local unitary transformations are
the unique family of states which show such a maximal
violation \cite{Chen}. Recently,  the  absolutely maximally entangled (AME) states have been defined and discussed \cite{Helwig, Raissi, limaosheng}. The characterization of AME states of an $n$-particle quantum system is that any collection of $\lfloor\frac{n}{2}\rfloor$ particles are in a maximally mixed state, where, $\lfloor\cdot\rfloor$  is the floor function.
Well-known examples are
the Bell and GHZ states on two and three parties respectively.
 Obviously, the 4-qubit GHZ state is not a absolutely maximally entangled state.

According to different  criteria \cite{1,2,3,4,5,6}, the maximally entangled states  can be divided into multiple types \cite{7,8,9,10,11}, such as the  AME state and planar maximally entangled (PME) state.  PME states have the property that any collection of  $\lfloor\frac{n}{2}\rfloor$ adjacent  particles in a planar $n$-particle quantum system are in a completely mixed state \cite{26}. Various maximally entangled states can be applied to quantum parallel teleportation \cite{Helwig,12,13,14,15}, quantum secret sharing \cite{16,17,18} and so on.

AME states do not exist in all possible Hilbert spaces .
It has been demonstrated that for any  $n$-particle system,  when the dimension $d$ of the Hilbert space of the subsystem  is chosen large enough, one can  find  the AME states; however, for low local dimensions there are  severe  constraints.
The existence of AME states composed of two-level systems was recently solved.
 For qubit systems ($d$=2),  AME states do only exist for $n=2,3,5$, and 6 parties, whereas it has been
shown that no AME states exist for 4, 7, 8, and more than
8 qubits \cite{Helwig,20,21,22,Higuchi,Scott, FHuber}.

Compared with AME state,  the PME state is one with a lower number of constraints.
Hence there are many more PME states than AME states,
 and  the PME states are  a wider class of entangled states than the AME states. The AME
states are a subclass of PME states, that is, any AME state is a PME state, but the converse is not true \cite{26}.  Is there other class of maximally entangled states except  the AME states and PME states? In this paper we solve this problem by  proposing a new class of maximally entangled states,  called the planar two-region multipartite maximally entangled states. In Sec. II, we investigate  a special case, the planar two-region four-partite maximally entangled states. Then in Sec. III, we study the exactly definition of this kind of maximally entangled states and prove that there exists this kind of maximally entangled state.  A summary with some remarks is provided in Sec. IV.

\section{Planar two-region four-partite maximally entangled state}

It was shown that there exist various PME states for any even number of qudits, and there are  two distinct multiparameter classes of four-qudit PMEs \cite{26}.
One may ask if it is possible to search for a wider class of entangled states except  AME states and PME states.

In this section we will answer the above problem by defining the planar two-region four-partite maximally entangled state first. Then we will show that there is this kind of maximally entangled state.

Assume that the structure of the planar quantum system consisting of  $n$ particles can be divided into two regions $A$, $B$, and each region has two disconnected parts. In the region $A$, one part has $k$ adjacent particles, and the other part  has $\lfloor n/2\rfloor-k$ adjacent particles.  At the same time we also require that one part of region $B$ has $k$ adjacent particles and the other part of the region $B$ has $n-\lfloor n/2\rfloor-k$ adjacent particles. As the two parts of region $A$  are not connected, so the only neighboring parts of each part of region $A$ are that of region $B$. Evidently, for a given planar quantum system there can exist many structures.  If a quantum state of the system satisfies that  every subset of region $A$ in each possible quantum structure, is in a completely mixed state, then  this state  is called a planar two-region four-partite maximally entangled state,  simply written as PKME$(n,k,4,d)$ state. Here $d$ is the dimension of the Hilbert space of single particle. Apparently, the requirement condition of this kind of maximally entangled state is weak than that of AME state and different from that of the planar maximally entangled state.

Next we will prove that there are PKME$(n,k,4,d)$ states for even $n$ and odd $n$, respectively.

Suppose that a partition of $n$ particles of a quantum system is parted  into parts $A$ and $B$ with $|A|\leq |B|$, then  a pure state $|\Psi\rangle$ of the quantum system can be written as \cite {26}
\begin{equation}
|\Psi\rangle=\sum_{\mathbf{k}}|\mathbf{k}\rangle_A|\phi_\mathbf{k}\rangle_B.
\end{equation}
So the density matrix of part $A$ is
\begin{equation}
\rho_A=\sum_{\mathbf{k,k'}}|\mathbf{k}\rangle_A\langle\mathbf{ k'}|\langle \phi_\mathbf{k}|\phi_{\mathbf{k'}}\rangle_B.
\end{equation}
For $\rho_A$ to be the completely  mixed state $I_A$, it is necessary to demand that the states $\{|\phi_{\mathbf{k}}\rangle\}$ are orthogonal and  with equal norm, and the states $\{|\mathbf{k}\rangle\}$ form a basis for the Hilbert space of the part $A$ \cite {26}.

\subsection{PKME$(n,k,4,d)$ states in a quantum system consisting of even number of particles}

\subsubsection{PKME$(4k,k,4,d)$ state in a 4$k$-qudit planar system}

 We start from a special case, in which  there are   $4k$ qudits  in a planar system, where $k$  is a positive integer. This planar $4k$-qudit system is depicted in  figure 1.

\begin{figure}[h]
\centering
\includegraphics[height = 39 mm,width=46 mm]{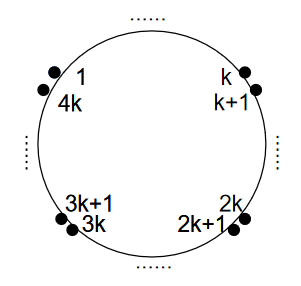}
\caption{The structure of the planar $4k$-qudit quantum system.
\label{FIG. 1.}}
\end{figure}

It is easy to see that there are $2k$ quantum structures  satisfying the requirements of the  PKME$(4k,k,4,d)$ state. That is, there exist $2k$ particle distributions satisfying that  region $A$ consists of two nonadjacent parts $A_{1}$ and $A_{2}$, region $B$ consists  of two nonadjacent parts $B_{1}$ and $B_{2}$, and each of  $A_{1}$, $A_{2}$, $B_{1}$ and $B_{2}$, has $k$ adjacent particles.

One can deduce that a  state in a Hilbert space $(C^{d})^{\bigotimes4k}$ of $4k$ qudits,
\begin{equation}
\begin{aligned}       
|\Psi\rangle=\frac{1}{d^{k}}\sum_{i_{1},i_{2},\cdots,i_{2k}=0}^{d}|i_{1},i_{2},\cdots,i_{k},i_{1},i_{2},\cdots,i_{k},i_{k+1},\cdots,i_{2k},i_{k+1},\cdots,i_{2k}\rangle
\end{aligned}
\end{equation}
is a PKME$(4k,k,4,d)$ state, where $C^d$ denotes a $d$-dimensional complex space.  Figure 2 shows the corresponding state structure.

\begin{figure}[h]
\centering
\includegraphics[height = 39 mm,width=46 mm]{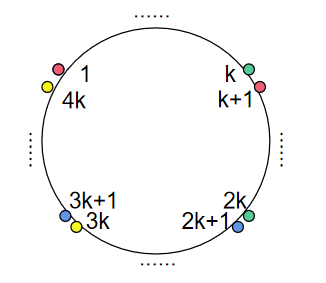}
\caption{A PKME($4k,k,4,d$) state of the planar $4k$-qudit  system, where both the 1st particle and the $(k+1)$th particle are  red,  $\cdots$, both the $k$th particle and the $2k$th particle are green, $\cdots$, both the $(2k+1)$th particle and the $(3k+1)$th particle are  blue, $\cdots$, and the $3k$th and $4k$th particles are both yellow. Here the same color represents the same state.
\label{FIG. 2.}}
\end{figure}

Therefore, there exists the PKME$(4k,k,4,d)$ state  of a planar $4k$-qudit  system.

\subsubsection{The family of PKME$(4,1,4,2)$ states of a 4-qubit planar system}

Let us consider a 4-qubit quantum system. The qubits are numbered as $\{1,2,3,4\}$ respectively. Clearly, there are two quantum structures in this system, which  are  $$\{A=\{A_1=1, A_2=3\},  B=\{B_1=2, B_2=4\}\},$$
$$\{A=\{A_1=2, A_2=4\},  B=\{B_1=1, B_2=3\}\},$$
and $k$ can only be 1. Next we will give the family of the PKME$(4,1,4,2)$ states.

The  state of the 4-qubit planar system can be expressed as
\begin{equation}
\begin{aligned}       
|\Psi\rangle=\frac{1}{2}\sum_{i,j,k,l=0}^{1}\alpha_{i,j,k,l}|i,j,k,l\rangle.
\end{aligned}
\end{equation}
Then it  can be written as
\begin{equation}
\begin{aligned}       
|\Psi\rangle_{13,24}=\frac{1}{2}(|00\rangle|Y_{00}\rangle+|01\rangle|Y_{01}\rangle+|10\rangle|Y_{10}\rangle+|11\rangle|Y_{11}\rangle)_{13,24},
\end{aligned}
\end{equation}
where
\begin{equation}
\begin{aligned}       
&|00\rangle|Y_{00}\rangle=\alpha_{0000}|0000\rangle+\alpha_{0001}|0001\rangle+\alpha_{0010}|0010\rangle+\alpha_{0011}|0011\rangle\text{,}\\
&|01\rangle|Y_{01}\rangle=\alpha_{0100}|0100\rangle+\alpha_{0101}|0101\rangle+\alpha_{0110}|0110\rangle+\alpha_{0111}|0111\rangle\text{,}\\
&|10\rangle|Y_{10}\rangle=\alpha_{1000}|1000\rangle+\alpha_{1001}|1001\rangle+\alpha_{1010}|1010\rangle+\alpha_{1011}|1011\rangle\text{,}\\
&|11\rangle|Y_{11}\rangle=\alpha_{1100}|1100\rangle+\alpha_{1101}|1101\rangle+\alpha_{1110}|1110\rangle+\alpha_{1111}|1111\rangle\text{.}\\
\end{aligned}
\end{equation}
According to the generalized Schmidt decomposition of multipartite quantum states \cite {Carteret}, one can always take
\begin{equation}
\alpha_{0001}=\alpha_{0010}=\alpha_{0100}=\alpha_{1000}=0
\end{equation}
by choosing a suitable basis.
Thus we have
\begin{equation}
\begin{aligned}       
&|Y_{00}\rangle=\alpha_{0000}|00\rangle+\alpha_{0011}|11\rangle\text{,}\\
&|Y_{01}\rangle=\alpha_{0101}|01\rangle+\alpha_{0110}|10\rangle+\alpha_{0111}|11\rangle\text{,}\\
&|Y_{10}\rangle=\alpha_{1001}|01\rangle+\alpha_{1010}|10\rangle+\alpha_{1011}|11\rangle\text{,}\\
&|Y_{11}\rangle=\alpha_{1100}|00\rangle+\alpha_{1101}|01\rangle+\alpha_{1110}|10\rangle+\alpha_{1111}|11\rangle\text{.}\\
\end{aligned}
\end{equation}

Obviously, if and only if  $\langle Y_{ik}|Y_{jl}\rangle=\delta_{ij}\delta_{kl}$, quantum state $|\Psi\rangle_{13,24}$ can satisfy $\rho_{13}=I$.  As a matter of fact $\langle Y_{ik}|Y_{jl}\rangle=\delta_{ij}\delta_{kl}$ implies that  the matrix
\begin{equation}       
V=\begin{pmatrix}  
    \alpha_{0000} & 0 & 0 & \alpha_{1100}\\  
    0 & \alpha_{0101} & \alpha_{1001} & \alpha_{1101}\\
    0 & \alpha_{0110} & \alpha_{1010} & \alpha_{1110}\\
    \alpha_{0011} & \alpha_{0111} & \alpha_{1011} & \alpha_{1111}\\
  \end{pmatrix}
\equiv \begin{pmatrix}  
    a & 0 & 0 & b\\  
    0 & c & e & f\\
    0 & g & h & p\\
    q & r & w & y\\
  \end{pmatrix}
\end{equation}
is a unitary matrix.

It is well-known that a unitary matrix has the properties: (a) Two different rows (columns) are orthogonal; (b) If there is an element whose magnitude is unit module, then all the other elements in the row and column are all 0 except for this element; (c) There is phase freedom globally.

Using the orthogonality of the first, second and third rows, one can obtain
\begin{equation}\label{10}
bf=bp=0.
\end{equation}
The solutions of above equation can be divided  the following two cases.

(i) $b = 0 $.  In this case, one must have   $|a|=1$ and $q=0$.  Without loss of generality, we can choose $a=1$,  which means $\alpha_{0000}=1$. Then the matrix $V$ becomes
\begin{equation}       
\begin{pmatrix}  
    1 & 0 & 0 & 0\\  
    0 & c & e & f\\
    0 & g & h & p\\
    0 & r & w & y\\
  \end{pmatrix},
\end{equation}
where the matrix $\begin{pmatrix}  
    c & e & f\\
    g & h & p\\
    r & w & y\\
  \end{pmatrix}$  is a $3\times 3$ unitary matrix.

Therefore  the quantum state can be expressed as
\begin{equation}
\begin{aligned}       
|\Psi'\rangle_{13,24}=&\frac{1}{2}(|0000\rangle+c|0101\rangle+g|0110\rangle+r|0111\rangle+e|1001\rangle+h|1010\rangle\\
&+w|1011\rangle+f|1101\rangle+p|1110\rangle+y|1111\rangle).
\end{aligned}
\end{equation}
It is easy to deduce that $\rho_{24}=I$. Hence, the quantum state $|\Psi'\rangle_{13,24}$ is   a family of PKME(4, 1, 4, 2) states of 4-qubit planar system.

(ii) $b\neq0$. In this case  we can obtain $f=p=0$ by Eq.(\ref{10}). Similarly,  we get $r=w=0$ by using the orthogonality of the first, second and third columns. Thus the matrix $V$ reads

\begin{equation}       
\begin{pmatrix}  
    a & 0 & 0 & b\\  
    0 & c & e & 0\\
    0 & g & h & 0\\
    q & 0 & 0 & y\\
  \end{pmatrix}.
\end{equation}
Evidently, if and only if the inner matrix $\begin{pmatrix}  
    c  & e\\  
       g &  h\
  \end{pmatrix}$ and outer matrix $\begin{pmatrix}  
    a  & b\\  
    q &  y\\
  \end{pmatrix}$ are $2\times2$  unitary matrices, then the matrix $V$ is a unitary matrix.

Hence in this case we have
\begin{equation}
\begin{aligned}       
|\Psi"\rangle_{13,24}=&\frac{1}{2}\{a|0000\rangle+q|0011\rangle
+b|1100\rangle+y|1111\rangle\\
&+c|0101\rangle+g|0110\rangle+e|1001\rangle+h|1010\rangle\}.
\\
\end{aligned}
\end{equation}
One can also derive that $\rho_{24}=I$. Therefore $|\Psi"\rangle_{13,24}$ is a PKME(4,1,4,2) state of a 4-qubit  planar system, if $\begin{pmatrix}  
    c  & e\\  
       g & h\\
  \end{pmatrix}$ and outer matrix $\begin{pmatrix}  
    a  & b\\  
    q &  y\\
  \end{pmatrix}$ are $2\times2$  unitary matrices.

 We use  $\{|\Psi_{0}\rangle\}$ to denote the intersection set of  $\{|\Psi'\rangle_{13,24}\}$ and $\{|\Psi"\rangle_{13,24}\}$. Obviously, one has
\begin{equation}
\begin{aligned}       
|\Psi_{0}\rangle_{13,24}&=\frac{1}{2}(|0000\rangle+c|0101\rangle+g|0110\rangle+e|1001\rangle+h|1010\rangle+|1111\rangle),
\end{aligned}
\end{equation}
where $\begin{pmatrix}  
    c  & e\\  
      g &  h\\
  \end{pmatrix}$ is a  $2\times2$  unitary matrix. Of course, the union   $\{|\Psi'\rangle_{13,24}\} \bigcup\{|\Psi"\rangle_{13,24}\}$ consists of all PKME(4, 1, 4,2) states of a 4-qubit planar system. So there are PKME(4,1,4,2) states of the 4-qubit planar system, although there is no AME state of 4-qubit system.

 Additionally, here  we provide an example of PKME state of   a 6-qubit planar system.  The qubits are denoted as 1,2,3,4,5,6 in a planar circle. Clearly,  there are 12 quantum structures
$$\{ A_1=\{1\},  A_2=\{3,4\}; B_1=\{2\},  B_2=\{5,6\}\},$$
$$\{ A_1=\{1\},  A_2=\{4,5\}; B_1=\{6\},  B_2=\{2,3\}\},$$
$$\{ A_1=\{2\},  A_2=\{4,5\}; B_1=\{3\},  B_2=\{1,6\}\},$$
$$\{ A_1=\{2\},  A_2=\{5,6\}; B_1=\{1\},  B_2=\{3,4\}\},$$
$$\{ A_1=\{3\},  A_2=\{5,6\}; B_1=\{4\},  B_2=\{1,2\}\},$$
$$\{ A_1=\{3\},  A_2=\{1,6\}; B_1=\{2\},  B_2=\{4,5\}\},$$
$$\{ A_1=\{4\},  A_2=\{1,6\}; B_1=\{5\},  B_2=\{2,3\}\},$$
$$\{ A_1=\{4\},  A_2=\{1,2\}; B_1=\{3\},  B_2=\{5,6\}\},$$
$$\{ A_1=\{5\},  A_2=\{1,2\}; B_1=\{6\},  B_2=\{3,4\}\},$$
$$\{ A_1=\{5\},  A_2=\{2,3\}; B_1=\{4\},  B_2=\{1,6\}\},$$
$$\{ A_1=\{6\},  A_2=\{2,3\}; B_1=\{1\},  B_2=\{4,5\}\},$$
$$\{ A_1=\{6\},  A_2=\{3,4\}; B_1=\{5\},  B_2=\{1,2\}\}.$$
One can easily check that a quantum  state in a Hilbert space $(C^{2})^{\bigotimes6}$
\begin{equation}
\begin{aligned}       
|\Psi\rangle=\frac{1}{2\sqrt{2}}\sum_{i,j,l=0}^{1}|i,i\oplus j,j,l,j\oplus l,i\oplus j\oplus l\rangle_{123456}\\
\end{aligned}
\end{equation}
is a PKME(6,1,4,2) state, where $k=1$.

\subsubsection{The constructed PKME states of the quantum systems with even number of particles}

Suppose that $|\Phi^{+}\rangle$  is a PKME$(4k, k, 4,d)$ state of a  $4k$-qudit quantum system. We  can construct a  more complicated PKME$(4k, k, 4,d)$ state by using the controlled  operators $\{\Lambda_ {s, t} (U)\}$. Here  $s$ and $t$ represent the control site   and target site respectively, $U$ is a unitary matrix acting on the $t$-th qudit \cite{30}. That is
\begin{equation}
\begin{aligned}
&\Lambda_{s,t}(U)|i\rangle_{s}|j\rangle_t\equiv |i\rangle_s|U(i,j)\rangle_t.\\
\end{aligned}\end{equation}
 Evidently,
\begin{equation}\label{18}
\begin{aligned}       
|\Psi\rangle=\Lambda_{k+1,k+2}(U_{1})\cdots\Lambda_{2k,3k+1}(U_{k})\cdots\Lambda_{4k-1,4k}(U_{2k-1})|\Phi^{+}\rangle
\end{aligned}
\end{equation}
is also a PKME$(4k,k,4,d)$ state of  a $4k$-qudit planar quantum system.

For example, from an 8-qudit PKME$(8,2,4,d)$ state
\begin{equation}\label{19}
\begin{aligned}       
|\Phi^{+}\rangle=\frac{1}{d^2}\sum_{i,j,l,m=0}^{d-1}|i,j,i,j,l,m,l,m\rangle_{1,2,3,4,5,6,7,8},
\end{aligned}
\end{equation}
one can make another PKME$(8,2,4,d)$ state
\begin{equation}\label{20}
\begin{aligned}       
|\Psi\rangle&=\Lambda_{3,4}(U_{1})\Lambda_{4,7}(U_{2})\Lambda_{7,8}(U_{3})|\Phi^{+}\rangle\\
&=\frac{1}{d^2}\sum_{i,j,l,m=0}^{d-1}|i,j,i,U_{1}(i,j),l,m,U_{2}(j,l),U_{3}(l,m)\rangle_{1,2,3,4,5,6,7,8},
\end{aligned}
\end{equation}
where states $|U(i,j)\rangle$ must satisfy $\langle U(i,j)|U(i,j')\rangle=\delta_{jj'}$. Figure 3 shows that when the input state is $|\Phi^{+}\rangle$ state, namely an 8-qudit PKME$(8,2,4,d)$ state, then the output state is also a PKME$(8,2,4,d)$.

\begin{figure}[h]
\centering
\includegraphics[height = 50 mm,width=80 mm]{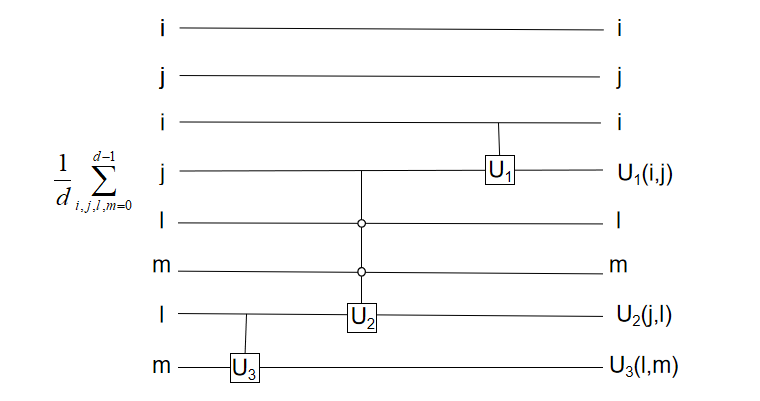}
\caption{The constructed PKME(8,2,4,$d$) state in an 8-qudit quantum system. The input state in the left is the quantum state in Eq.(\ref{19}), and the output state in the right is the quantum state in Eq.(\ref{20}).
\label{FIG. 3.}}
\end{figure}

By changing the order of action of the controlled operator $\Lambda_{s,t}(U)$ in Eq.(\ref{18}), we  can construct another  PKME$(8,2,4,d)$ state
\begin{equation}\label{21}
\begin{aligned}       
|\Psi\rangle=\Lambda_{4k-1,4k}(U_{2k-1})\cdots\Lambda_{2k,3k+1}(U_{k})\cdots\Lambda_{k+1,k+2}(U_{1})|\Phi^{+}\rangle.
\end{aligned}
\end{equation}
The   difference between  Eq. (\ref{18}) and Eq. (\ref{21}) is the order of action of the controlled operators $\Lambda_{s,t}(U)$. By using state in Eq. (\ref{19}), a new  PKME$(8,2,4,d)$ state of an 8-qudit planar quantum system is
\begin{equation}
\begin{aligned}       
|\Psi\rangle&=\Lambda_{7,8}(U_{3})\Lambda_{4,7}(U_{2})\Lambda_{3,4}(U_{1})|\Phi^{+}\rangle\\
&=\frac{1}{d^2}\sum_{i,j,l,m=0}^{d-1}|i,j,i,U_{1}(i,j),l,m,U_{2}(U_{1}(i,j),l),U_{3}(U_{2}(U_{1}(i,j),l),m)\rangle_{1,2,3,4,5,6,7,8},\\
\end{aligned}
\end{equation}
which is illustrated in figure 4.
\begin{figure}[h]
\centering
\includegraphics[height = 50 mm,width=80 mm]{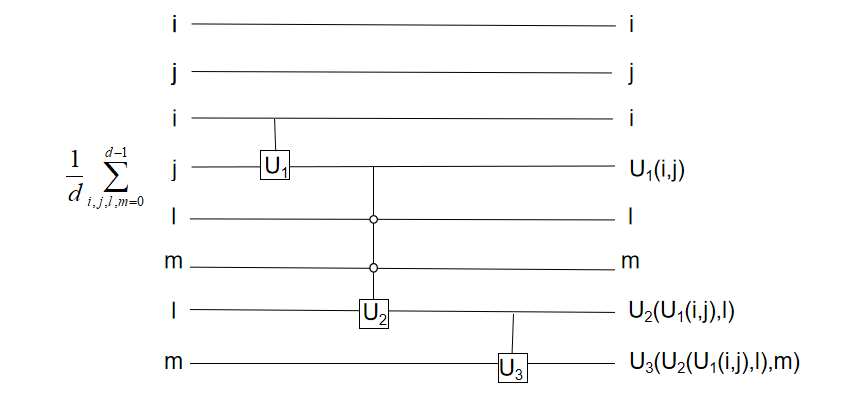}
\caption{A new PKME$(8,2,4,d)$ state of an 8-qudit quantum system.
\label{FIG. 4.}}
\end{figure}

\subsubsection{PKME$(4k+1,k,4,2)$ state of the $4k+1$  particle system}

Let us first consider  a special case $k=1$, i.e., a 5-qubit quantum system. Five qubits in a planar circle system are numbered as 1,2,3,4,5, respectively. It is not difficult to check that the quantum state
\begin{equation}
\begin{aligned}       
|\Psi\rangle=\frac{1}{2}\sum_{i,j=0}^{1}|i,i,j,j,i\oplus j\rangle_{1,2,3,4,5}
\end{aligned}
\end{equation}
is a PKME$(5,1,4,2)$ state. Here there are 5 quantum structures
$$\{ A_1=\{1\},  A_2=\{3\}; B_1=\{2\},  B_2=\{4,5\}\},$$
$$\{ A_1=\{1\},  A_2=\{4\}; B_1=\{5\},  B_2=\{2,3\}\},$$
$$\{ A_1=\{2\},  A_2=\{4\}; B_1=\{3\},  B_2=\{1,5\}\},$$
$$\{ A_1=\{2\},  A_2=\{5\}; B_1=\{1\},  B_2=\{3,4\}\},$$
$$\{ A_1=\{3\},  A_2=\{5\}; B_1=\{4\},  B_2=\{1,2\}\}.$$
Therefore, there  exist PKME$(n,k,4,d)$  states of the planar  quantum system with odd-particles.

\begin{figure}[h]
\centering
\includegraphics[height = 42 mm,width=46 mm]{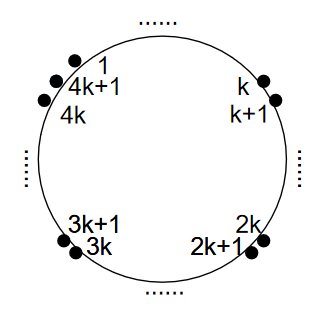}
\caption{Planar $4k+1$-qubit quantum system.
\label{FIG. 5.}}
\end{figure}

Now we consider the generalized case,  a $(4k+1)$-qubit planar  system in a circle, where $k\geq 1$ is an arbitrary  positive integer.
Each quantum structure consists of four parts $\{A_{1}, A_{2}, B_{1}, B_2\}$.  Here   $A_{1}$, $A_{2}$, $B_{1}$, and  $B_{2}$ contain  $k$, $k$, $k$, and $k+1$ neighboring particles, respectively;  region $A$ is composed of parts $A_{1}$ and  $A_{2}$, region $B$ is composed of parts $B_{1}$ and  $B_{2}$, as shown in figure \ref{FIG. 5.}.  Clearly, there are $4k+1$ quantum structures in this planar  system.

\begin{figure}[h]
\centering
\includegraphics[height = 46 mm,width=46 mm]{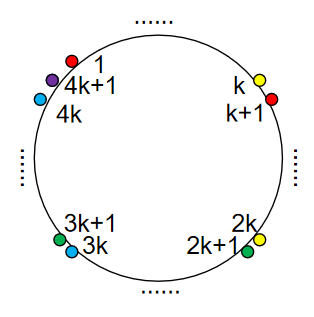}
\caption{The quantum state of the planar $(4k+1)$-qubit quantum system in Eq.(\ref{26}). The 1st particle and the $k+1$th particle are red, the $k$th particle and the $2k$th particle are yellow, the $2k+1$th particle and the $3k+1$th particle are green, the $3k$th particle and the $4k$th particle are blue, and the $(4k+1)$th particle is purple. Here different colors represent different states.
\label{FIG. 6.}}
\end{figure}

A quantum  state in a Hilbert space $(C^{2})^{\bigotimes4k+1}$ of the $4k+1$ qubits
\begin{equation}\label{26}
\begin{aligned}       
|\Psi\rangle=&\frac{1}{2^{k}}\sum_{i_{1},i_{2},\cdots,i_{2k}=0}^{1}|i_1\rangle_1|i_2\rangle_2\cdots|i_k\rangle_k|i_1\rangle_{k+1}|i_2\rangle_{k+2}\cdots|i_k\rangle_{2k}\\
&|i_{k+1}\rangle_{2k+1}|i_{k+2}\rangle_{2k+2}\cdots|i_{2k}\rangle_{3k}|i_{k+1}\rangle_{3k+1}|i_{k+2}\rangle_{3k+2}\cdots|i_{2k}\rangle_{4k}|i_{1}\oplus i_{2}\oplus\cdots \oplus i_{2k}\rangle_{4k+1}
\end{aligned}
\end{equation}
is a PKME$(4k+1, k, 4,2)$ state, which  is plotted in figure \ref{FIG. 6.}.

\subsubsection{Constructed PKME$(4k+1, k, 4, d)$ state of a   quantum system with odd number of particles}

Suppose that $|\Phi^{+}\rangle$ state is a PKME$(4k+1, k, 4, d)$ state of a $(4k+1)$-qudit quantum system. One can generate  a new PKME$(4k+1, k, 4, d)$ state
\begin{equation}\label{27}
\begin{aligned}       
|\Psi\rangle=\Lambda_{k+1,k+2}(U_{1})\cdots\Lambda_{2k,3k+1}(U_{k})\cdots\Lambda_{4k,4k+1}(U_{2k})|\Phi^{+}\rangle.
\end{aligned}
\end{equation}

\begin{figure}[h]
\centering
\includegraphics[height = 50 mm,width=70 mm]{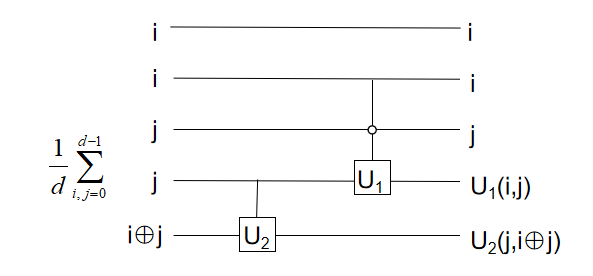}
\caption{The constructed PKME state in a 5-qubit quantum system. The input state in the left is a state (\ref{28}), and the output state in the right is the  state (\ref{29}). Here $d=2$.
\label{FIG. 7.}}
\end{figure}

For example, one can use a 5-qubit PKME$(5, 1, 4, d)$ state
\begin{equation}\label{28}
\begin{aligned}       
|\Phi^{+}\rangle=\frac{1}{d}\sum_{i,j=0}^{d-1}|i,i,j,j,i\oplus j\rangle_{1,2,3,4,5}
\end{aligned}
\end{equation}
to construct another PKME$(5, 1, 4, d)$ state
\begin{equation}\label{29}
\begin{aligned}       
|\Psi\rangle&=\Lambda_{2,4}(U_{1})\Lambda_{4,5}(U_{2})|\Phi^{+}\rangle\\
&=\frac{1}{d}\sum_{i,j=0}^{d-1}|i,i,j,U_{1}(i,j),U_{2}(j,i\oplus j)\rangle_{1,2,3,4,5},
\end{aligned}
\end{equation}
as depicted in figure \ref{FIG. 7.}, where $d=2$ and the $|U(i,j)\rangle$ state satisfies $\langle U(i,j)|U(i,j')\rangle=\delta_{ii}\delta_{jj'}=\delta_{jj'}$.

\begin{figure}[h]
\centering
\includegraphics[height = 50 mm,width=70 mm]{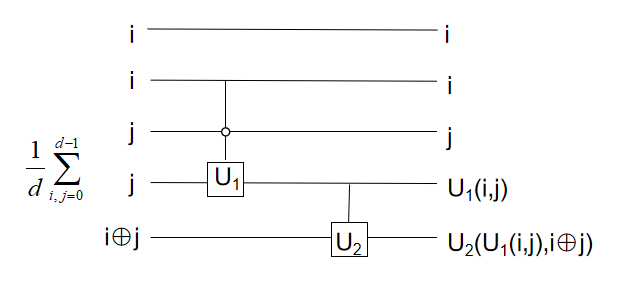}
\caption{Another constructed PKME state of a 5-qudit quantum system.  Here $d=2$, and the input state in the left is the quantum state  (\ref{28}), and the output state in the right is the state (\ref{31}).
\label{FIG.8.}}
\end{figure}

By changing the order of action of the controlled operators $\{\Lambda_{s,t}(U)\}$ in Eq.(\ref{27}) we can construct another  PKME$(4k+1,k, 4, d)$ state
\begin{equation}
\begin{aligned}       
|\Psi\rangle=\Lambda_{4k,4k+1}(U_{2k})\cdots\Lambda_{2k,3k+1}(U_{k})\cdots\Lambda_{k+1,k+2}(U_{1})|\Phi^{+}\rangle.
\end{aligned}
\end{equation}
The difference between this state and the state stated by  Eq.(\ref{27}) is only the order of action of the controlled operator $\{\Lambda_{s,t}(U)\}$. Taking state in Eq.(\ref{28}) as an example, the new  PKME$(5,1,4,d)$ state of this  5-qubit system reads
\begin{equation}\label{31}
\begin{aligned}       
|\Psi\rangle&=\Lambda_{4,5}(U_{2})\Lambda_{2,4}(U_{1})|\Phi^{+}\rangle\\
&=\frac{1}{d}\sum_{i,j=0}^{d-1}|i,i,j,U_{1}(i,j),U_{2}(U_{1}(i,j),i\oplus j)\rangle_{1,2,3,4,5},
\end{aligned}
\end{equation}
which is illustrated in figure \ref{FIG.8.}. Here $d=2$.

\subsubsection{PKME(7,2,4,2) state }

Let us consider a 7-qubit quantum system. There are 7 quantum structures
$$\{ A_1=\{1,2\},  A_2=\{5\}; B_1=\{3,4\},  B_2=\{6,7\}\},$$
$$\{ A_1=\{2,3\},  A_2=\{6\}; B_1=\{4,5\},  B_2=\{1,7\}\},$$
$$\{ A_1=\{3,4\},  A_2=\{7\}; B_1=\{5,6\},  B_2=\{1,2\}\},$$
$$\{ A_1=\{4,5\},  A_2=\{1\}; B_1=\{6,7\},  B_2=\{2,3\}\},$$
$$\{ A_1=\{5,6\},  A_2=\{2\}; B_1=\{1,7\},  B_2=\{3,4\}\},$$
$$\{ A_1=\{6,7\},  A_2=\{3\}; B_1=\{1,2\},  B_2=\{4,5\}\},$$
$$\{ A_1=\{1,7\},  A_2=\{4\}; B_1=\{2,3\},  B_2=\{5,6\}\}.$$
It is easy to prove that  a quantum state in a Hilbert space $(C^{2})^{\bigotimes7}$ of 7-qubit system
\begin{equation}
\begin{aligned}       
|\Psi\rangle=\frac{1}{2\sqrt{2}}\sum_{i,j,l=0}^{1}|i,j,l,j,l,i,i\oplus j\oplus l\rangle_{1,2,3,4,5,6,7}
\end{aligned}
\end{equation}
is a PKME(7,2,4,2) state. However,  AME  state of seven qubits do not exist.

\section{Planar two-region multi-partite maximally entangled state}

 In this section, we will generalize the definition of two-region four-partite  maximally entangled state to the more generalized case. That is we will  define two-region multi-partite  maximally entangled state  and prove that  this kind of maximally entangled states do exist.

  Consider a quantum system composed of  $n$ particles in a planar circle, where the Hilbert space of each particle  is a $d$-dimensional complex space $C^d$. Suppose that one can divide the $n$ particles into two-region $\{A,B\}$, where $A$ ($B$) consists of parts $\{A_1, A_2,\cdots, A_m\}$ ($\{B_1, B_2,\cdots, B_m\}$), and the particles in each part are adjacent, however $A_1, A_2,\cdots, A_m$ are not neighbors, so do $B_1, B_2,\cdots, B_m$.   Each $\{A_1, A_2,\cdots, A_m\, B_1, B_2,\cdots, B_m\}$ is called a quantum structure. We also demand that region $A$ contains $\lfloor n/2\rfloor$ particles, region $B$ consists of  $n-\lfloor n/2\rfloor$ particles. We use  $k_{A_1}, k_{A_2}, \cdots, k_{A_m}, k_{B_1}, k_{B_2}, \cdots, k_{B_m}$ to denote  the corresponding  particle numbers of parts $A_1, A_2,\cdots, A_m, B_1, B_2,\cdots, B_m$ respectively. Here $2m\leq n,  4\leq n$, $|A|\leq |B|$. Obviously, for an  $n$ particle system  and fixed   $k_{A_1}, k_{A_2}, \cdots, k_{A_m}, k_{B_1}, k_{B_2}, \cdots, k_{B_m}$ there are many quantum structures.

   If a quantum state in the Hilbert space $(C^{d})^{\bigotimes n}$ of the $n-$particle system satisfies that  every subset of region $A$ in each possible quantum structure, is in a completely mixed state, then  this state  is called a planar two-region multi-partite maximally entangled state,  simply written as PKME$(n,\{z\},2m,d)$ state. Here $d$ is the dimension of the Hilbert space of single particle and $\{z\}=\{k_{A_1}, k_{A_2}, \cdots, k_{A_m}, k_{B_1}, k_{B_2}, \cdots, k_{B_m}\}$.

Clearly, the requirement condition of PKME states  is weak than that  of the absolutely maximally entangled state and different from that of  the planar maximally entangled state.

Next we will give two examples  of  the planar two-region multi-partite maximally entangled states.

{\it Example 1}.  Consider a $2mk$-qubit system.  It is easy to demonstrate that the quantum  state in Hilbert space $(C^{2})^{\bigotimes2mk}$
\begin{equation}
\begin{aligned}       
|\Psi\rangle=&\frac{1}{2^{\frac{mk}{2}}}\sum_{i_{1},i_{2},\ldots,i_{mk}=0}^{1}|i_{1}\rangle_1\cdots|i_{k}\rangle_k|i_{1}\rangle_{k+1}
\cdots|i_{k}\rangle_{2k}|i_{k+1}\rangle_{2k+1}\cdots|i_{2k}\rangle_{3k}|i_{k+1}\rangle_{3k+1}\cdots|i_{2k}\rangle_{4k}\cdots\\
&|i_{(m-1)k+1}\rangle_{(2m-2)k+1}\cdots|i_{mk}\rangle_{(2m-1)k}|i_{(m-1)k+1}\rangle_{(2m-1)k+1}\cdots|i_{mk}\rangle_{2mk}
\end{aligned}
\end{equation}
is a PKME$(2mk,\{z\}, m, 2)$ state, with $\{z\}=\{k_{A_1}, k_{A_2}, \cdots, k_{A_m}, k_{B_1}, k_{B_2}, \cdots, k_{B_m}|k_{A_1}=k_{A_2}=\cdots=k_{A_m}=k_{B_1}=k_{B_2}= k_{B_m}=k \}$.

Therefore, there exists the planar two-region multi-partite maximally entangled state of  the planar $2mk$-qubit system, where both $k$ and $m$ are positive integers larger than or equal to 1.

{\it Example 2}. Assume $m$, $k$ are positive integers larger than or equal to 1. Let us consider a $2mk+1$ qubit system. One can verify that a  quantum state in the Hilbert space  $(C^{2})^{\bigotimes2mk+1}$
\begin{equation}
\begin{aligned}       
|\Psi\rangle=&\frac{1}{2^{\frac{mk}{2}}}\sum_{i_{1},i_{2},\ldots,i_{mk}=0}^{1}|i_{1}\rangle_1\cdots|i_{k}\rangle_k|i_{1}\rangle_{k+1}
\cdots|i_{k}\rangle_{2k}|i_{k+1}\rangle_{2k+1}\cdots|i_{2k}\rangle_{3k}|i_{k+1}\rangle_{3k+1}\cdots|i_{2k}\rangle_{4k}\cdots\\
&|i_{(m-1)k+1}\rangle_{(2m-2)k+1}\cdots|i_{mk}\rangle_{(2m-1)k}|i_{(m-1)k+1}\rangle_{(2m-1)k+1}\cdots|i_{mk}\rangle_{2mk}|\oplus_{j=1}^{mk} i_j\rangle_{2mk+1}
\end{aligned}
\end{equation}
is just a  PKME$(2mk+1,\{z\}, m, 2)$ state, with $\{z\}=\{k_{A_1}, k_{A_2}, \cdots, k_{A_m}, k_{B_1}, k_{B_2}, \cdots, k_{B_m}|k_{A_1}=k_{A_2}=\cdots=k_{A_m}=k_{B_1}=k_{B_2}= k_{B_{m-1}}=k, k_{B_{m}}=k+1 \}$.

Thus, there is the planar two-region multi-partite maximally entangled state  of the planar $2mk+1$-qubit  system, where both $k$ and $m$ are positive integers larger than or equal to 1.

\section{Summary}

In this paper, we  discuss a new kind of  maximally entangled states, the planar two-region  multi-partite maximally entangled states.  It is  demonstrated that  there are the planar two-region four-partite maximally entangled states in 4-qubit and 7-qubit planar systems, although there is no AME state in 4-qubit and 7-qubit systems. Furthermore we  show that there are the planar two-region four-partite maximally entangled states whether in the quantum system with even particles or in that with odd particles. Additionally, based on some  planar two-region four-partite maximally entangled states, we also construct the new  planar two-region four-partite maximally entangled states. Some important examples of  the planar two-region four-partite and multi-partite maximally entangled states are provided.

Evidently, the virtue  of two-region multi-partite maximally entangled states may induce many applications, especially  in quantum parallel teleportation and quantum secret sharing. One can easily design the protocols of quantum parallel teleportation and quantum secret sharing by using the similar method provided by the authors of Ref. \cite{26}. We hope these protocols can be realized in the future experiments.

\begin{acknowledgments}
This work was supported by the National Natural Science Foundation of China under Grant Nos. 62271189, 12071110,  and the Hebei Central Guidance on Local Science and Technology Development Foundation of China under Grant No. 236Z7604G.
\end{acknowledgments}

\end{document}